\documentclass[prb,aps,floatfix,amsmath,twocolumn]{revtex4}

\usepackage{amsmath}
\usepackage{epsfig}
\usepackage{array}
\usepackage{verbatim}
\usepackage{hyperref}
\usepackage{amssymb}
\usepackage{multirow}
\usepackage{graphicx}

\usepackage{color}
\usepackage{ulem}

\newcommand{\re}[1]{(\ref{#1})}

\newcommand{\beg}{\begin{equation}}

\newcommand{\en}{\end{equation}}

\newcommand {\dis}{\displaystyle}

\newcommand{\eps}{\varepsilon}
\newcommand{\lam}{\lambda}
\newcommand{\nnn}{\nonumber}

\newcommand{\eref}[1]{Eq.~(\ref{#1})}

\renewcommand{\emph}{\textit}

\newcommand{\beq}{\begin{equation}}

\newcommand{\eeq}{\end{equation}}
\newcommand{\barray}{\begin{eqnarray}}
\newcommand{\earray}{\end{eqnarray}}

\newcommand{\disp}[1]{Eq.~(\ref{#1})}

\newcommand{\ve}{\varepsilon}

\newcommand{\tp}{t^{+}}
\newcommand{\tm}{t^{-}}
\newcommand{\iden}{{\bf 1}}
\newcommand{\PM}{\Pi}
\newcommand{\SM}{{\cal S}}

\begin{document}

\title{Quantum integrability in systems with finite number of levels}

\author{Emil A. Yuzbashyan$^1$} 
\author{B. Sriram Shastry$^2$}

\affiliation{$^1$Center for Materials Theory, Department of Physics and Astronomy, Rutgers University, Piscataway, NJ 08854, USA\\
$^2$Physics Department, University of California, Santa Cruz, CA 95064, USA}

\begin{abstract}

We consider the problem of defining quantum integrability in systems with finite number of energy levels starting from commuting matrices and construct new general classes of such matrix models with a given number of commuting partners. We argue that  if the matrices depend on a (real) parameter, one can  define quantum integrability from this feature alone, leading  to specific results such as exact solvability, Poissonian energy  level statistics and  to level crossings.
 
 \end{abstract}
 
\date{\today}
\maketitle

\section{Introduction}

The field  of quantum integrable systems, originally a somewhat abstruse  topic,  has received a great deal of attention  in recent years with the realization that many typical models are realizable in atomic, mesoscopic  and  macroscopic  many-body systems. In  describing molecular systems such as benzene,  we  deal with Hamiltonians defined on a finite-dimensional Hilbert space. These arise from  a lattice  of finite   size,    with  spins, fermions or bosons populating the sites.  In condensed matter systems too, one often studies finite size systems as a prelude to taking the thermodynamic limit
or to obtain finite size corrections. Our concern in this work is with he construction of general quantum integrable models, their characterization and definition of quantum integrability in  a finite dimensional  context\cite{comment-finite}, a task that is considerably more  delicate  than the corresponding classical case.

 We may view   typical condensed matter problems   as  models with a finite number of discrete single-particle energy levels, and  sometimes (but not always) with a  fixed number of particles. The Hamiltonian  is then `just' an $N\times N$ Hermitian matrix, with a suitable $N$. This description is, by design, far removed from its  parentage  in the space of  many body models.   Given such a Hamiltonian matrix, can we say whether it is integrable or not? What is the precise notion of quantum integrability in this case; can we separate $N\times N$ Hermitian matrices into two clearly distinct classes -- integrable and otherwise? If yes, what are the consequences of quantum integrability that can be \textit{derived from its definition}, i.e. the characteristic properties of  such `integrable' matrices?

In this paper, we present a coherent  view point that emerges from our recent  exploration of  such questions. We present a summary of our previous\cite{yuzbashyan,shastry,owusu,shastry1,owusu1} and new results and list some open questions that remain unanswered. To put this enquiry into context, note  that in the thermodynamic limit  $N\to \infty$,  impressive alternative approaches   are available. For example Baxter's work on the spectra of infinite dimensional corner transfer matrices\cite{baxter-ctm} culminates in the realization of  extra symmetries and  structure arising in that limit, formalized by the Yangian approach\cite{yangian}. Another  viewpoint is of geometric origin due to Sutherland\cite{sutherland} and emphasizes non-diffractive scattering as a true hallmark of quantum integrability. Much as we admire these powerful  viewpoints, our chosen task of understanding finite dimensional matrix systems takes us in quite a different direction.

In classical mechanics a system with $n$ degrees of freedom is said to be integrable if it has $n$ {\it functionally independent} integrals of motion  that Poisson commute\cite{arnold}, i.e. are in involution. An exact solution of equations of motion follows from this definition, and it further can be shown that the motion is confined to invariant tori cut out in the phase space by the conservation laws. Unfortunately, a straightforward import of this elegant notion of integrability into quantum mechanics is problematic. 

The  difficulties are at least two fold: firstly what is a degree of freedom? In quantum models with a fixed number of particles, this number seems closest to the classical notion of a degree of freedom.  However,   in addition to the particle number there are other integer parameters to take into account such as the number of sites or the magnitude of spins. For example,  in the Hubbard model with $L$ sites do we count the number of electrons $n_e$ or the number of holes $2L-n_e$?  Or does the Heisenberg model for $L$ spin-$\frac{1}{2}$ particles have the same number of degrees of freedom as that for $L$ spin-$5$ particles?

The second  difficulty is in defining what constitutes a nontrivial integral of motion. From an elementary theorem in algebra\cite{perlis}, we know that an arbitrary $N \times N$  Hermitian matrix commutes with $N$ other Hermitian matrices that may be chosen as the projection operators along each component in the diagonal representation. We could equivalently express the matrix and all its commuting partners as a power series in some chosen non degenerate matrix.  A similar statement can be proved for general commuting Hermitian operators with discrete or continuous spectra \cite{neumann,neumann1,weigert}.
 If now we are given a special Hamiltonian matrix, and told that  it is quantum integrable (in some specific sense that is yet to be defined),  possessing $n$  equally special commuting partner matrices;  we might  (rather stubbornly) ignore this extra information and  compute the $N$   elementary commuting partners  of the given matrix as per the earlier prescription. Don't we then  have too many partners?  What,  if any,  is the distinction between these two sets of commuting partners? From such questions and  considerations one seems to be forced to call integrable either all or none Hamiltonians with finite Hilbert spaces.

Our goal here is to attempt a resolution of these difficulties and provide a practically useful, simple and at the same time  rigorous definition of quantum integrability for Hamiltonian matrices. Our interest is of course not  in the definition per se but  mostly in  a systematic construction of new quantum integrable models and a clear delineation and derivation of their properties starting from the definition.  Characteristics normally attributed to integrable models include: exact solution for their spectra, Poisson level statistics, crossings of levels of the same symmetry in parameter-dependent Hamiltonians etc.  
However, they are never derived from one another or within some unified framework, but each property has to be established independently and on a model by model basis.

Consider, for example, the 1d Hubbard model. Shortly after the model was exactly solved via Bethe's Ansatz\cite{lieb}, a separate rigorous study by Heilmann and Lieb\cite{heilman} of its parameter independent symmetries  revealed numerous level crossings  violating  the Wigner von Neuman non-crossing rule for eigenvalues in generic systems. These findings can be rationalized by the   notion  that  copious   level crossings are  a hallmark of quantum integrabilty, and hint at parameter dependent conservation laws in such a model.
Furthermore, by solving the Bethe ansatz equations\cite{deguchi}, Êit was shown  that some degenerate energy levels that can not be explained in terms of ÊU-independent symmetries have distinct values with respect to higher conserved quantities.
 Much later, one of us\cite{S,shastry-hubbard} found an embedding of the 1d Hubbard model into the Yang--Baxter scheme, thereby displaying an explicit set of parameter dependent conservation laws. Not surprisingly, a numerical 
analysis\cite{poilblanc} of level spacings finds Poisson statistics that crosses over to GOE when `integrability' is destroyed by adding an off-site Coulomb repulsion. Further 
work\cite{yuzbashyan} by us on making a causal link between the conservation laws and the level crossings in the 1d Hubbard model has been a fruitful source of insights that we have subsequently explored  and report here.   
  
Often, one of the above characteristics is singled out and adopted as a definition (see e.g. Ref.~\onlinecite{caux} for examples). 
It seems to us that while level statistics and crossings are useful, and even powerful  tests of integrability, these  are hardly suitable as {\it definitions}. These criteria  arguably encompass a broader class of systems than those normally thought  integrable \cite{localization}.  It is also difficult to see how, starting from either  of them, one could derive other properties, e.g. an exact solution. 
Universal statistics emerges only in the limit of large matrices, $N\to\infty$, or for an ensemble of matrices, while here we are looking for a notion that also works for a stand-alone Hamiltonian matrix with fixed $N$. In addition, there are  exceptional points in parameter space of systems usually recognized as integrable, where the statistics is non-Poissonian, e.g. for Gaudin magnets\cite{relano}.   Starting from a well-defined  notion of integrability that we propose below, we construct  broad classes of new integrable models where  such deviations occur more generally   and  explain their origin. Similarly, we find examples that fail the level crossing test -- such as `accidental' degeneracies in non-integrable systems and (rare) instances of integrable systems without crossings.

It turns out that such a simple and yet well-defined notion can in fact be formulated as we recognized in a series of papers\cite{yuzbashyan,shastry,owusu,shastry1,owusu1}.  The main distinguishing feature of our approach is that it  leads in a unified fashion to a general construction of new   quantum integrable models, their exact solution as well as allows for a systematic study of various properties such as level crossings, level statistics etc. The main idea is to consider the dependence of  commuting operators on a real parameter, which we denote $u$. This is typically an interaction constant or an applied field, e.g. Coulomb interaction constant in the Hubbard model,  magnetic field in Gaudin magnets\cite{sklyanin},  pairing strength in the  BCS and anisotropy in the $XXZ$ Heisenberg models,  etc.   Besides usual space-time and internal space symmetries, which are parameter independent, these models have  $u$-dependent conservation laws (alternatively termed dynamical symmetries or conserved currents) as discussed above.    For the BCS and Gaudin models  the Hamiltonian and all conserved currents are linear in the parameter\cite{sklyanin,cambiaggio}. In the $XXZ$ and Hubbard models the Hamiltonian and the first dynamical symmetry are linear in $u$; the rest are polynomials in it of order two and higher\cite{S,Lu,Gr,GM,links,zhou,Fu}.  

 The rest of the paper is organized as follows. In Sects.~\ref{def} through \ref{lcross} we present a comprehensive summary of our previous results. For a detailed discussion and relevant derivations we refer the reader to 
Refs.~\onlinecite{shastry,owusu,shastry1,owusu1}. Sects.~\ref{yb} and \ref{lstat} are devoted to new results. In Sect.~\ref{yb} we provide a link between our notion of quantum integrability and the conventional approach based on the Yang--Baxter equation. Sect.~\ref{lstat} summarizes our upcoming publication\cite{hansen} on level statistics in models that are integrable under our definition.

\section{\label{def}Definition}

In  several examples of  quantum integrable models of interest, such as the anisotropic Heisenberg model and the Hubbard model,  there is a set of commuting operators linear  or polynomial in
an interaction type parameter, termed $u$ here. This set of course includes the Hamiltonian, which is typically linear in $u$- e.g. the interaction constant $U$ for the Hubbard model.  This seems to be the most common  situation in parameter dependent integrable models and leads us\cite{shastry,owusu,shastry1,owusu1} to consider operators of the form $H(u)=T+uV$, where $T$ and $V$ are $N\times N$ Hermitian matrices. A key observation is that fixed $u$-dependence implies a natural  well-defined notion of a nontrivial integral of motion. A typical (e.g. randomly generated) $H(u)$  commutes only with $(a+bu)\iden + c H(u)$, a trivial operator linear in $u$. Here $a, b$, and $c$ are real numbers and $\iden$ is the identity operator. 

The requirement that there exist a nontrivial commuting partner linear in $u$  severely constrains the matrix elements of $H(u)$. In fact, as we will see below,  a real symmetric $H(u)$ that has such a partner is fixed by  less than $(N-1)(N+8)/2$ real parameters. In contrast, $N(N+1)$ real parameters are necessary  to specify the matrix elements of $T$ and $V$ for a generic $H(u)$, indicating that matrices with fixed parameter-dependence split into two non-overlapping sets -- those with nontrivial commuting partners and those without. 

Moreover, there is a natural classification of integrable families according to the number $n$ of linearly independent   commuting operators. Specifically, we define an \textit{integrable family} as a vector space of $n$ linearly independent $N\times N$ Hermitian matrices 
\begin{eqnarray}
H^k(u)& =& T^k+uV^k \;\mbox{such that,}  \nnn \\
\ [H^i(u),H^j(u)]& =&0\,\,\mbox{for all $u$ and $i, j=1,\dots,n$}.
\label{comm}
\end{eqnarray}
In addition, we impose an (optional) condition  that $H^i(u)$ have no common $u$-independent symmetries -- there is  no  constant matrix $\Omega$ ($\neq a\iden$) such that $ [\Omega,H^i(u)]=0$ for all $u$ and $i$. If there are such symmetries, $H^i(u)$ are simultaneously block-diagonal and \eref{comm} reduces to that for smaller matrices (blocks) without $u$-independent symmetries\cite{hubnote}. 

 Linear independence means that $\sum_i a_i H^i(u)=(a+bu)\iden$ with real $a_i$  if and only if all $a_i=0$ and $a=b=0$. Note that, for convenience, we chose to separate multiples of identity from our list of nontrivial commuting matrices $H^k(u)$.   Finally, $n$ is defined as the maximum number of nontrivial independent commuting matrices in a given family, i.e. any $H(u)=T+u V$ that commutes with all $H^i(u)$ can be written as 
\beg
H(u)=\sum_{i=1}^{n} a_k H^k(u)+(a+bu)\iden.
\label{vecspace}
\en
 Thus $H^i(u)$ act as  basis vectors in the $n$-dimensional vector space  ${\cal V}_n$ of commuting matrices, defined up to a multiple of identity matrix $\iden$. 

We propose the following classification of integrable matrices linear in a parameter into 
\textit{types}.
The maximum possible number of linearly independent $N\times N$ commuting Hermitian matrices is $n=N-1$, not counting multiples of the  identity matrix. We call this a  \textit{type 1} or maximally commuting family of matrices. Similarly,  families with $n=N-2$ independent commuting $H^i(u)$ are termed \textit{type 2}, and a general \textit{type M} family is defined through $n=N-M$. The maximum value of $M$ is $M=N-2$ when there are \textit{two} nontrivial members in the family 
\textit{in addition} to  $(a+bu)\iden$. Note that $M=N-1$ means $n=1$, i.e. an arbitrary \textit{nonintegrable} matrix that has no commuting partners  besides itself and the identity.  Since we study integrable matrices in this paper, everywhere below $M\ge N-2$. 

We argue that the following properties are   direct  consequences of the above definition of quantum integrability: (1) an exact solution for the eigenspectra of $H^i(u)$ in terms of roots of a \textit{single} algebraic equation, (2) $H^i(u)$ satisfy Yang--Baxter equation, (3) eigenvalues of $H^i(u)$ typically (but not always) cross as functions of $u$; the number of crossings depends on both $N$ and $M$ (4) eigenvalues have Poisson statistics in $N\to\infty$ limit except for some  special cases of certain measure zero in the space of all integrable families. Below we \textit{prove}  statements (1) through (3) for type 1 families (in this case crossings are always present) and  comment on similar results of \cite{owusu1} for other types.  Poisson statistics will be demonstrated numerically in a separate paper; here we briefly discuss some of its main findings.

\eref{comm} can be solved for matrix elements of $H^i(u)$ at least for some types of integrable families. All type 1 families were constructed in \cite{shastry,owusu}. In \cite{owusu1} all type 2, 3 and some type $M$ for arbitrary $M$ were obtained. To facilitate further discussion, let us first cast \eref{comm} into a different form. Using $H^k(u)=T^k+uV^k$ and equating to zero terms at all orders of $u$, we obtain
\beg
[T^i, V^j]=[T^j,V^i],\quad [T^i, T^j]=[V^i, V^j]=0.
\label{comm1}
\en
It is convenient to choose the basis in the target Hilbert space to be the common eigenbasis of the  mutually commuting matrices $V^i$. The first commutation relation in \eref{comm1} in this basis reads \cite{deg} $T^i_{km}/(d^i_k-d^i_m)=T^j_{km}/(d^j_k-d^j_m)\equiv S_{km}$, where $d^i_k$ are the diagonal elements of $V^i$. This implies that $T^i$ can be written as
\beg
T^i = W^i+ \left[ V^i,S \right]
\label{T}
\en
where $W^i$ is a diagonal matrix. Note that the antihermitian matrix $S$ is the same for all members of the family and is therefore independent of the  basis in ${\cal V}_n$. Now the commutation relations 
$[V^i,V^j]=0$ and $[T^i, V^j]=[T^j,V^i]$ are satisfied. The remaining equation, $[T^i, T^j]=0$, takes the form
\beg
\left[ [V^i,S], [V^j,S] \right]=\left[[V^j,S],W^i \right]- \left[[V^i,S],W^j \right].
\label{comm2}
\en

Before we proceed with  type 1, let us discuss the number of  parameters involved in constructing a generic real symmetric integrable family.

We can generate integrable families of various types by solving \eref{comm2} numerically\cite{owusu1}. The algorithm is as follows. First, we arbitrarily specify some diagonal matrices $V^{i}, V^{j}, W^{i}$ and $W^j$ ($4N$ real inputs) and solve for the antisymmetric matrix $S$. This yields a discrete set of solutions for $S$ and two commuting matrices $H^{i,j}(u)=W^{i,j}+[V^{i,j}, S]+ u V^{i,j}$ for each $S$. To determine, the remaining basic matrices $H^k(u)$ in the family with a given $S$, we take $j=k$ in 
\eref{comm2} (with $V^i$, $W^i$ and $S$  obtained before) and solve this equation for $V^k$ and $W^k$. 

Some choices of the $4N$ inputs in the above procedure produce the same commuting family. In fact, as seen from 
\eref{vecspace}, there is  $2(n+2)$-parametric freedom (values of $a_k$, $a$ and $b$) in picking two matrices within the family. This means that we can fix $2(n+2)$ out of $4N$ parameters  by taking linear combinations within the family, i.e. by going to a different basis in the vector space of commuting matrices. In addition, Eqs.~\re{comm2} and \re{T} are invariant with respect to rescaling $V^i\to\alpha V^i$ and
$S\to  S/\alpha$, which fixes one more parameter in $V^i$. This leaves $4N-2(n+2)-1=2N+2M-5$ real parameters to specify a generic type $M$ family.

To further select a particular matrix $H(u)=T+uV$ within the family, one needs to  pick $n+2$ coefficients in \eref{vecspace}, in addition to $2N+M-5$ parameters that specify the commuting family, i.e. a general type $M$ matrix involves $4N-n-3=3N+M-3$ real parameters. Since 
$M\le N-2$ the maximum number is $4N-5$. This counting has been done in the common eigenbasis of 
$V^i$. Going to an arbitrary basis in the target Hilbert space adds another $N(N-1)/2$ real parameters for a general orthogonal transformation. We see that the total number of parameters is less than $(N+8)(N-1)/2$, about half of $N(N+1)$ for a non-integrable real-symmetric matrix of the form $A+uB$. 

\section{\label{typeone}Type 1}

The `master' equation \re{comm2} is simplest for type 1. In this case there are $N$ linearly independent $H^i(u)$ (including the identity) and consequently $N$ linearly independent diagonal matrices $V^i$. By taking linear combinations  we can go to a  basis in the vector space ${\cal V}_{N-1}$ such that $V^i_{kk}\equiv d_k=\delta_{ik}$. \eref{comm2} reads $f^i_{jk}\equiv W^i_{jj}- W^i_{kk}=-S_{ij} S_{ik}/S_{jk}$.  These equations are consistent when $f^i_{jk}+f^i_{kl}+f^i_{lj}=0$, yielding the following four index relation\cite{shastry}:
\beg
 S_{ij} S_{jk} S_{kl} S_{li}+S_{ik} S_{kl} S_{lj} S_{ji}+S_{il} S_{lj} S_{jk} S_{ki}=0.
\label{nu}
\en
The most general solution of this equation is \cite{owusu,shastry1}
\beg
S_{jk}=\dfrac{\gamma_j \gamma_k^*}{\varepsilon_j-\varepsilon_k},
\label{S1}
\en
where real $\eps_i$ and complex $\gamma_i$ are unrestricted parameters that fix the commuting family. 

Next, using \eref{nu} and $V^i_{kj}=\delta_{kj}\delta_{ik}$, we determine $W^i$ from \eref{comm2}, which is linear in $W^i$, and $T^i$ from \eref{T}. The most general member of a type 1 family, $H(u)=\sum_{i=1}^N d_i H^i(u)$ with arbitrary real $d_i$, is
\begin{equation}
\begin{array}{l}
\dis \left[H \left(u\right)\right]_{mn}= \gamma_{m}\gamma_{n}^*
\left(\frac{d_{m} -d_{n} }{\varepsilon_{m}-\varepsilon_{n}}\right),\quad m\ne n,\\
\\
\dis \left[H \left(u\right)\right]_{mm}=u\, d_{m} -\sum_{k\neq
m}|\gamma_{k}|^{2}
\left(\frac{d_{m} -d_{k} }{\varepsilon_{m}-\varepsilon_{k}}\right).\\
\end{array}
\label{H}
\end{equation}
Note that $d_m$ are eigenvalues of $V$ by design. In particular, for basic operators $H^i(u)$ we have $d_k=\delta_{ik}$, i.e.
their nonzero matrix elements are
\begin{equation}
\begin{array}{l}
\dis \left[H^i \left(u\right)\right]_{ij}= \frac{\gamma_{i}\gamma_{j}^*}{\varepsilon_{i}-\varepsilon_{j} },\quad j\ne i,\\
\\
\dis \left[H^i \left(u\right)\right]_{jj}=u\, \delta_{ij} -
\sum_{k\neq j}|\gamma_{k}|^{2}
\left(\frac{\delta_{ij} -\delta_{ik} }{\varepsilon_{m}-\varepsilon_{k}}\right).\\
\end{array}
\label{Hi}
\end{equation}

\section{\label{typeM}Type $M>1$}

A similar construction is possible for real symmetric integrable families of arbitrary type $M$\cite{owusu1}, though the expressions for the matrix elements are somewhat more involved. Specifically, we have (see Ref.~\onlinecite{owusu1} for the derivation and more details)
\beg
S_{kl}=\dfrac{1}{2} \dfrac{\gamma_k \gamma_l}{\varepsilon_k-\varepsilon_l}\left( \Gamma_k+\Gamma_l \right),
\label{full form}
\en
where 
\beg
\Gamma_m\equiv\Gamma(\eps_m),\quad \Gamma(\sigma)   =\pm \sqrt{\dfrac{\prod_{j=1}^{M}{(\phi_j-\sigma)}}{\prod_{j=1}^{M}{(\lambda_j-\sigma)}}}, 
\label{gams}
\en
$\gamma_i, \eps_i$ are  arbitrary real parameters playing the same role as in Type 1. The sign in \eref{gams} can be chosen at will individually for each $\eps_m$. In addition, type $M$ features $M$ new real parameters $\phi_i$ that need to be chosen so that the radicand in \eref{gams} is real (see below). 

Finally, quantities $\lambda_i$  in \eref{gams} are by construction solutions of the following equation with arbitrary real $B$:
\beg
f(\lam_i)\equiv\sum_{j=1}^{N}\dfrac{\gamma_{j}^2}{\lambda_i-\varepsilon_j}=B.
\label{Gaudin1}
\en
All $N$ roots $\lam_i$ of this equation are real.  Indeed,  $f(\lam)\to +\infty$ as $\lam\to \eps_k^+$
and $f(\lam)\to -\infty$ as $\lam\to \eps_{k+1}^-$, where $\eps_k$ are ordered, 
$\eps_1<\eps_2<\dots<\eps_N$. It follows that  $f(\lam)=B$ has a real solution between $\eps_k$ and
$\eps_{k+1}$ for any $k$, i.e. $\eps_m<\lam_m<\eps_{m+1}$.  
One more root is located above $\eps_N$, $\lam_N>\eps_N$, for $B>0$ and below $\eps_1$, $\lambda_1<\eps_1$, for $B<0$, where we ordered $\lam_k$ so that $\lam_1<\lam_2<\dots<\lam_N$. 

To ensure the reality of $\Gamma(\eps_m)$ for any $\eps_m$ it is sufficient (though not necessary) to choose parameters $\phi_i$ so that $\eps_i< \phi_i<\lam_i$  for $B>0$ and 
$\lambda_i < \phi_i < \varepsilon_i$  for $B<0$. This is the only restriction on $\phi_i$.

The most general member of this type $M$ commuting family is
$$
\begin{array}{l}
\dis \left[H \left(u\right)\right]_{mn}=
\gamma_{m}\gamma_{n} \left(\frac{d_{m} -d_{n} }{\varepsilon_{m}-\varepsilon_{n}}\right) \dfrac{\Gamma_m +\Gamma_n}{2},\quad m\ne n,\\
\\
\dis\left[H \left(u\right)\right]_{mm}=u\, d_{m} -\\
\\
\dis \qquad \sum_{j\neq m} {\gamma_{j}^{2}
\left(\frac{d_{m} -d_{j} }{\varepsilon_{m}-\varepsilon_{j}}\right) \dfrac{1}{2} \dfrac{\left( \Gamma_m+\Gamma_j\right)\left( \Gamma_j+1\right)}{\Gamma_m+1}},\\
\end{array}
\label{Hmel2}
$$
where, unlike the Type 1 case, $d_m$ are not arbitrary, but are given by
\beg
d_m=g_0+\sum_{j=1}^{N-M}\frac{g_j}{\lambda_{j+M}-\varepsilon_m}
\label{dm2}
\en
and $g_j$ are arbitrary real numbers.  

Commuting families obtained by the above prescription (termed \textit{ansatz type $M$} families in Ref.~\onlinecite{owusu1}) contain $2N+M+1$  arbitrary parameters   -- $2N$ of  $\gamma_i$'s and $\varepsilon_i$'s, $M$ of$P_i$'s and the parameter $B$. As discussed in detail in Ref.~\onlinecite{owusu1}, there are certain gauge transformations, such as a uniform scaling of $\gamma_i$ and $\varepsilon_i$ or a uniform shift of $\varepsilon_i$,  that leave the commuting family invariant.  This allows to fix three   of the parameters meaning that the number of parameters needed to uniquely specify a Type $M$ commuting family produced by this construction is $2N+M-2$.  

On the other hand, we argued in Sect.~\ref{def} based on numerical evidence and other considerations that a {\it general} Type $M \geq 3$ family is uniquely specified by $2N+2M-5$ parameters. This suggests that our construction  can produce 
\textit{all} real symmetric commuting families
only for $M=1,2,3$, while for $M>3$ it yields only a subset of such families. The completeness for $M=1,2$ was explicitly demonstrated in Ref.~\onlinecite{owusu,owusu1}, while for Type $M=3$ it is supported by numerical tests.

In fact, a correspondence between  real symmetric type $M$ families and  compact Riemann surfaces of genus $g\ge M-1$ was conjectured in 
Ref.~\onlinecite{owusu1}. It turns out that  the above formulas  produce families that correspond to hyperelliptic  Riemann surfaces of genus $g=M-1$. All  Riemann surfaces of genus 0, 1, and 2 are hyperelliptic which explains the completeness of the construction for $M=1,2,3$.

\section{\label{exsol}Exact solution}

The exact spectra of type 1 matrices were obtained in \cite{owusu}. The components of an
eigenvector (column) $\vec{v}_m (u)$ of $H(u)$ given by \re{H} are
\beg
  \left[ \vec{v}_m (u) \right]_j= \dfrac{\gamma_j}{\lambda_m-\varepsilon_j},
\label{eigenvector1}
\en
with respective eigenvalue
\beg
E_m(u)=\sum_{k=1}^N{\dfrac{d_k|\gamma_k|^2}{\lambda_m-\varepsilon_k}},
\label{eH}
\en
where the $\lambda_i$, $i=1, \dots, N$ are determined from a single algebraic equation
\beg
f(\lam_m)\equiv \sum_{j=1}^N{\dfrac{|\gamma_j|^2}{\lambda_m-\varepsilon_j}}=u.
 \label{constraint1}
\en
 As discussed in the previous section all $N$ roots $\lam_i$ of this equation are real.

The above equations can be verified directly by evaluating $\sum_j [H(u)]_{ij} [ \vec{v}_m (u) ]_j$. 
Ansatz type $M$    families of Sect.~\ref{typeone} have an exact solution in terms of a single equation similar to \eref{constraint1}
\beg
 \sum_{j=1}^N{\dfrac{1}{2} \dfrac{\gamma_j^2}{\sigma-\varepsilon_j} \left(\Gamma(\sigma) + \Gamma_j \right)}-\dfrac{B}{2}\left( \Gamma(\sigma)-1\right)=u.
 \label{constraint}
\en

Having solved this equation for $\sigma$, we obtain the eigenvalues
\beg
E_\sigma(u)=\sum_{k,j}{\dfrac{d_k}{\lambda_k-\varepsilon_j} \dfrac{\gamma_j^2}{\sigma-\varepsilon_j} \dfrac{1}{2} \big(\Gamma_j+\Gamma(\sigma) \big)},
\label{eigened}
\en
and the corresponding eigenvectors
 \beg
  \left[\vec{v}_\sigma (u)\right]_j= \dfrac{1}{2} \dfrac{\gamma_j}{\sigma-\varepsilon_j} \left( \Gamma(\sigma) + \Gamma_j\right).
\label{eigenvector}
\en

\section{\label{lcross}Level crossings}

First, we show that any type 1 matrix  $H(u)=T+uV$ has at least one level crossing\cite{owusu}. To this end, let us analyze the evolution of eigenvalues $E_m(u)$ with $u$. We  observe    that
$\lam_m\to \eps_m$ as $u\to+\infty$. In this limit the main contribution to  Eqs.~\re{eH} and 
\re{constraint1} comes from the $j=m$ term, yielding   $E_m \to |u| d_m$.
Similarly,  we obtain $x_m\to \eps_{m+1}$ and $E_m\to -|u|d_{m+1}$ for $u\to-\infty$. It is not surprising that eigenvalues of $H(u)$ tend to $\pm |u|d_k$ since $d_k$ are eigenvalues of $V$ and $uV$ dominates $H(u)$ for large $u$. What is important however is that we know to which particular $ud_k$ a given $E_m(u)$ tends in both limits. Symbolically, we can write $k\to k-1 \mbox{ (mod $N$)}$ meaning the eigenvalue  goes from $-|u|d_k$ on the left ($u\to -\infty$)   to $|u|d_{k-1}$ on the right ($u\to +\infty$).

The presence of levels crossings can now be proved by contradiction. Suppose there are no crossings. Since eigenvalues are continuous functions of $u$, this implies that their ordering must be the same at all $u$. The top level must connect the largest eigenvalue at $u\to -\infty$ to the largest eigenvalue at $u\to +\infty$, the bottom level goes from the lowest eigenvalue at 
$u\to -\infty$ to that at $u\to +\infty$, etc. Let $d_k$ be ordered as\cite{degenerate} $d_i<d_j<\dots<d_m$. Then, the largest (lowest) eigenvalue at $u\to -\infty$ is $-|u| d_i$ ($-|u| d_m$) and 
the largest (lowest) eigenvalue at $u\to +\infty$ is $|u|d_m$ ($|u|d_i$), i.e. we have $i\to m$ and $m\to i$ for the top and bottom levels, respectively. On the other hand, according to the $k\to k-1$  
rule established above
this   implies $m=i-1\mbox{ (mod $N$)}$ and at the same time $i=m-1\mbox{ (mod $N$)}$. We obtain $0=2\mbox{ (mod $N$)}$,
which does not hold for any $N\ge 3$, i.e. the above assumption that
levels do not cross cannot be true. Thus, at least one level crossing is inevitable.

\begin{figure}[t!]
\begin{centering}
\includegraphics[scale=0.2]{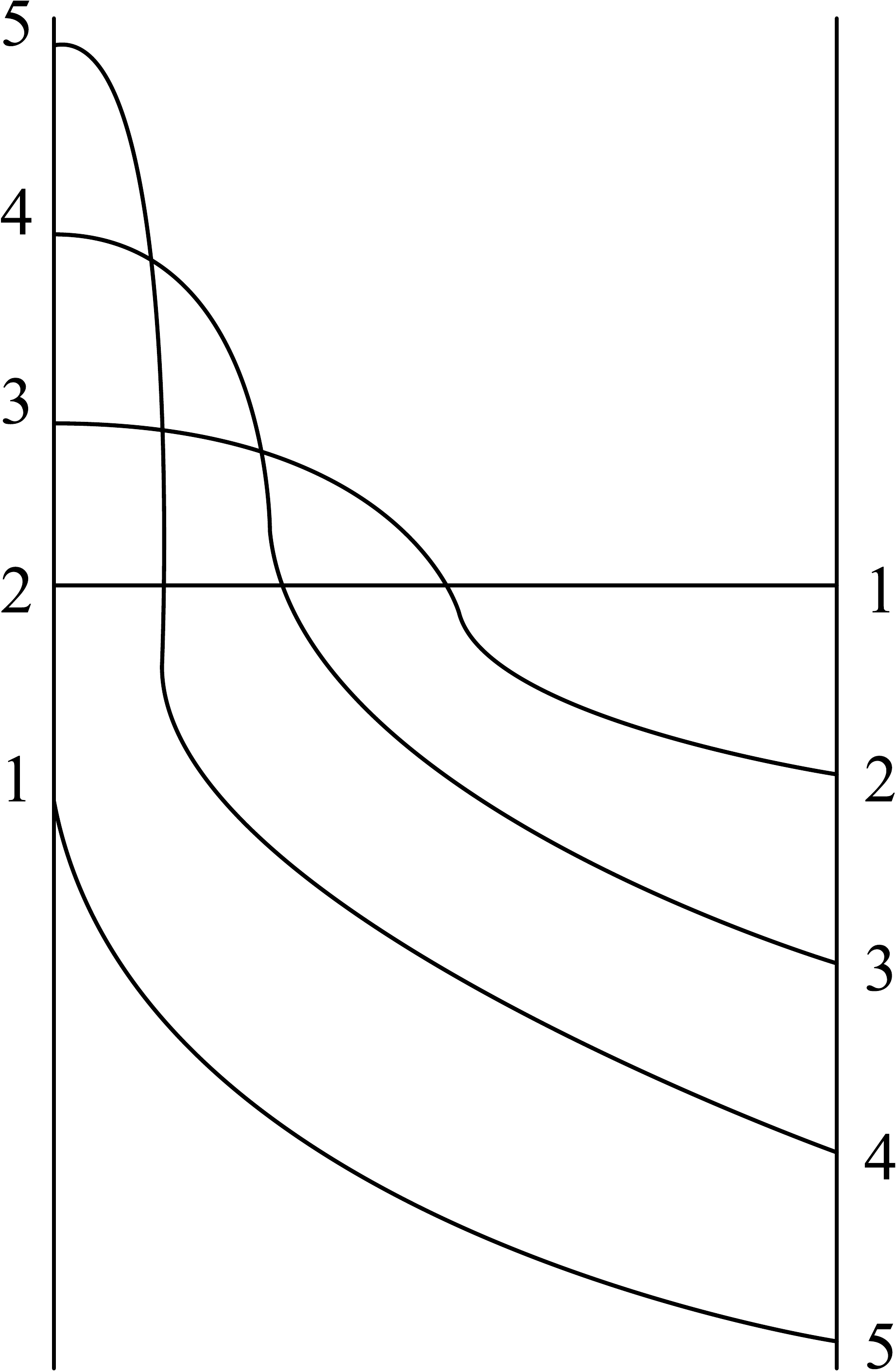} 
\end{centering}
\caption{\label{maxcross}  A schematic energy level diagram for $N=5$ demonstrating maximum number of crossings $(N-1)(N-2)/2$. The diagram is for the ordering $d_1>d_2>\dots >d_N$ and uses the rule $k\to k-1$ derived in the text.
 }
\end{figure}

 Allowed values of the total number of crossings $n_\times$ in a type 1 matrix can be determined by analyzing the representation of an arbitrary type 1 matrix $\widetilde{H}(u)$ in terms of powers of any other nontrivial $H(u)$ that belongs to the same commuting family. One finds\cite{owusu} that this expansion is necessarily of the form
\beg
\widetilde{H}(u)=\sum_{m=0}^{N-1}{\dfrac{Q^m(u)}{P^H(u)} \left[ H(u) \right]^m},
\label{expand}
\en
where $P^H(u)$ is a polynomial in $u$ of degree $(N-1)(N-2)/2$ with real coefficients that depend on matrix elements of $H(u)$ only. $Q^m(u)$ are polynomials in $u$ of order $m-1$ lower than $P^H(u)$.
This expansion breaks down at a given value of $u=u_\times$ only  when $H(u)$ has a crossing at $u=u_\times$ and  $P^H(u_\times)=0$. Thus, crossings of $H(u)$ occur at the roots of $P^H(u)$
(see \cite{owusu} for a detailed proof). We conclude that the maximum number of crossings in a type 1 matrix is 
\beg
n_\times^{\max}=\frac{(N-1)(N-2)}{2}.
\label{max}
\en
This upper bound is realized e.g. for matrices
\re{H} such that $d_1>d_2>\dots >d_N$, see Fig.~\ref{maxcross}. $n_\times$ has a definite parity, that of $n_\times^{\max}$, because the coefficients of $P^H(u)$  are real and its complex roots therefore come in  conjugate pairs. For example, $4\times 4$ type 1 matrices have either 1 or 3 crossings, $n_\times=2,4,6$ for $5\times 5$ etc.

Crossings are also ubiquitous in type $M>1$ integrable families. It is possible however to deliberately engineer  $4\times 4$ type 2 matrices \textit{without} crossings\cite{owusu}. Other than that,
we do not have rigorous results for higher types. Empirically, one finds\cite{owusu1} that the expansion \re{expand} still holds, but the degree of $P^H(u)$ and therefore the maximum number of crossings is reduced to $(N-1)(N-2)/2-g$, where $g\ge M-1$ is the genus of the corresponding Riemann surface, typically $g=M-1$ (see the end of Sect.~\ref{typeM}).

Interestingly, one can also construct $4\times 4$ real symmetric matrices $A+uB$ that have crossings but no $u$-independent symmetry and no commuting partners linear in $u$ (this is impossible for $N=3$), see \cite{yuzbashyan,owusu} for details. This indicates that either `accidental' degeneracies are possible or one can meaningfully introduce nonlinear in $u$ conserved currents in such cases.

\section{\label{yb}Yang--Baxter equation for the Matrix Model}

We next show how the Type-1 matrices can be fit into the Yang--Baxter formulation of integrable systems, by displaying matrix objects $S$, \disp{smatrix-m}, that play the role of scattering amplitudes\cite{yang}.  Let us note that this construction has not yet been realized for general Type $M$ matrices. Thus we consider  a purely matrix model in $N$ dimensions with states $|i\rangle$,  $1\leq i \leq N$,  the projection operators $ \pi_{ij}$  and the identity matrix $\iden$:
$$
\begin{array}{l}
\dis \pi_{ij} = |i\rangle \langle j|  \\
\\
\dis \iden = \sum_i \pi_{ii}\\
\end{array}
$$
With  $i \neq j$, let us define  a `dressed' permutation operator
$$
\PM_{ij}= \frac{|\gamma_i|^2+|\gamma_j|^2}{2}  \iden+ \gamma_i \gamma_j^*  ( \pi_{ij}  + \pi_{ji}  ) -|\gamma_i|^2  \pi_{jj} - |\gamma_j|^2  \pi_{ii}.
$$
If we set $\gamma_i \to \gamma_j$,  
  $\PM$ reduces to the permutation operator
 \beq
 \PM_{ij} \to |\gamma_j|^2 \ P_{ij},
 \eeq
 where $P_{ij}$   acts as: 
 \beq 
 P_{ij} |k \rangle = \delta_{ik} |j \rangle + \delta_{jk} |i \rangle + (1-\delta_{ik})(1-  \delta_{jk}) |k \rangle. \eeq
 
 Let us introduce $g$ as a coupling parameter and also  the composite parameter $x=(\ve,\gamma)$ so that the scattering operator $\SM_{ij}$ is defined as:

\beq
\SM_{ij}(x_j|x_i) \equiv  \SM_{ij}= \frac{(\ve_j-\ve_i) \iden+ 2g \ \PM_{ij}}{(\ve_j-\ve_i)+ g \  (|\gamma_i|^2+|\gamma_j|^2)} \label{smatrix-m}
\eeq

The action of a particular scattering matrix $\SM_{nm} $  on the relevant states is given by 
\barray
\SM_{n m} \  |k \rangle &=&  |k \rangle, \;\; \;\; (k \neq n,m) \nnn \\
\SM_{n m} \   \ | n \rangle &=& \tp_{nm} \   \ |n \rangle + r_{nm} \  \ |m\rangle \nnn \\
\SM_{n m} \  \ |m \rangle &=&\tm_{nm} \    \ |m \rangle +  r_{nm} \   \ |n\rangle  \label{s-matrix}
\earray
with the reflection ($r$) and transmission ($t$) amplitudes  defined by

\barray
r_{nm} &= & \frac{ {2g}  \gamma_n \gamma_m^*}{\ve_m-\ve_n + g( |\gamma_m|^2+|\gamma_n|^2)}, \nnn \\
\tp_{nm} &=& \frac{\ve_m-\ve_n + {g}( |\gamma_n|^2-|\gamma_m|^2)}{\ve_m-\ve_n + {g} ( |\gamma_m|^2+|\gamma_n|^2)}, \nnn \\
\tm_{nm} &= &   \frac{\ve_m-\ve_n + {g} ( |\gamma_m|^2-|\gamma_n|^2)}{\ve_m-\ve_n + {g} ( |\gamma_m|^2+|\gamma_n|^2)}. \label{rts}
\earray

  We   state   the two operator relations that are needed, and readily verified, for any three indices $1,2,3$:
 \barray
 \PM_{12} \ \PM_{32} \ \PM_{31} &=& \PM_{31} \  \PM_{32} \  \PM_{12},  \label{threep-m} \\
~[  \PM_{12},  \  ( \PM_{31}+ \PM_{32})] &=& 0. \label{permreln-m}
 \earray

With $C_{ij}$ as constants  we can now verify the condition
\barray
&&(C_{12} \iden+ g \ \PM_{12}) \ (C_{32} \iden+ g \ \PM_{32}) \ (C_{31} \iden+ g \ \PM_{31}) = \nnn \\
&& (C_{31} \iden+ g \ \PM_{31}) \ (C_{32}\iden+ g \ \PM_{32}) \ (C_{12}\iden+ g \  \PM_{12}),
\earray
provided $C's$ satisfy the triangle law:
\beq
C_{32}= C_{31}+ C_{12}.
\eeq
Dividing by a suitable constant and consulting \disp{s-matrix} we  therefore verify the Yang--Baxter equation for the $S$
\beq
\SM_{i k} \SM_{jk} \SM_{ij} = \SM_{ij} \SM_{jk} \SM_{i k}, \label{yb-matrix}
\eeq
and the initializing  condition
\beq
\lim_{x_j \to x_i} \SM_{ij}(x_j|x_i) \to P_{ij}.
\eeq

We consider the $N$ sites (indices) and add an auxiliary  index $\alpha$ (that is outside the original space of $N$ states)  so that
 $ \pi_{i \alpha} , \pi_{\alpha i} , \pi_{\alpha \alpha}$  are  added to the list of operators. Now define a (monodromy) matrix
\beq
{\cal T}_{\alpha} = B_\alpha (x_\alpha) \ \SM_{ N \alpha} \ \SM_{N-1 \ \alpha} \cdots \SM_{1 \alpha}.
\eeq 
 This is a function of all the  indicated variables:
\beq
{\cal T}_{\alpha} \equiv {\cal T}_{\alpha}(x_\alpha| \left\{x_1,x_2, \ldots, x_N \right\}),
\eeq
and also a  boundary field term  $  B_\alpha (x_\alpha)$. For the boundary term (twist) to give commuting operators, this  term must be chosen to satisfy the condition\cite{boundary}
\beq
~[\SM_{\alpha \beta}, B_\alpha B_\beta]=0,
\eeq
whereby we choose
\beq
B_\alpha(x_\alpha)  = \iden + \frac{g}{u} \   \  \pi_{\alpha \alpha}, \label{boundary-m}
\eeq
with the freedom of an  arbitrary parameter $u$.

 The transfer matrix  is obtained by tracing over the auxiliary index  $\alpha$ 
\beq
{\bf T}(x_\alpha |  \left\{x_1,x_2, \ldots, x_N \right\})  = \sum_\alpha \langle \alpha  |  {\cal T}_{\alpha} |\alpha \rangle.
\eeq
Using   Baxter's  classic proof \cite{baxter} for commutation of transfer matrices, slightly generalized to the case of twisted  boundary conditions \cite{boundary,sklyanin},  we conclude that
\beq
[{\bf T}(x_\alpha ), {\bf T}(x_\beta)]=0,\label{commute-m}
\eeq
for arbitrary $x_\alpha$ and $x_\beta$ while holding  $\{ x_j\} \, \ g, \ u$ fixed.  
We note that \disp{commute-m} is  also valid as  $x_\alpha \to x_i $  and hence conclude $ [{\bf T}_i,{\bf T}_j]=0 $, where 
\barray
{\bf T}_j &=& \lim_{x_\alpha\to x_j} {\bf T}(x_\alpha |  \left\{x_1,x_2, \ldots, x_N \right\}) \nnn\\
&=& \SM_{j-1,j}(x_j,x_{j-1}) \ldots  \SM_{1,j}(x_j,x_{1}). \nnn \\
&&B_{j}(x_j)   \SM_{N,j}(x_j,x_{N})\ldots  \SM_{j+1,j}(x_j,x_{j+1}).
\earray
An expansion in powers of the interaction strength $g$ produces the currents:
\beq
{\bf T}_j= \iden + \frac{g}{u}  H^j(u) + O(g^2), 
\eeq
with
$$
H^i(u) = \pi_{ii} + u  \sum_j \frac{ \gamma_i \gamma_j^* (\pi_{ij}+\pi_{ji} ) -|\gamma_i|^2 \pi_{jj} - |\gamma_j|^2 \pi_{ii}}{\ve_i-\ve_j}. 
$$
Considering terms of order $O(g)$ in $ [{\bf T}_i,{\bf T}_j]$,  we conclude 
\beq
[H^i(u), H^j(u)]=0. \label{commuting-z}
\eeq
Note that $H^i(u)$ are the  exactly the basis operators of Eq.~\re{Hi} written in terms of projection operators\cite{shastry1}.

\section{Links to various models}

First, note that one can choose an arbitrary Hermitian matrix $V$ and still have $H(u)=T+uV$ to be a member of any given type 1 family. Indeed, \eref{H} is written in the eigenbasis of $V$. Both this basis and the eigenvalues of $V$ -- parameters $d_k$ -- can be chosen arbitrarily.  By symmetry one can instead choose   $T$ at will, though this is not apparent from \eref{H}. But as soon as e.g. $V$   is fixed, $T$   is severely constrained -- one is left with only $2N$ parameters $\gamma_i, \eps_i$ to specify its matrix elements. This ability to choose either $V$ \textit{or} $T$ arbitrarily means in particular that any $u$-independent Hamiltonian can be `embedded' into a type 1 family in many different ways. For example, one can choose $V$ to be the isotropic Heisenberg   model, or the  Haldane-Shastry model, and find $T$ so that $T+uV$ belongs to a given type 1 family.

Type 1 integrable families are closely related\cite{owusu}  to Gaudin magnets\cite{sklyanin}
$\hat h_i=B\hat s_i^z +
\sum_{j\ne i} \hat {\bf s}_i \hat {\bf s}_j(\eps_i-\eps_j)^{-1}$, where $\hat {\bf s}_j$ are quantum spins of arbitrary length $s_j$. In the sector with (conserved)  
$\hat S^z=\sum_j \hat s_i^z$ equal to its maximum (minimum) possible value less (plus) one,
$\hat h_i$ are $N$ commuting $N\times N$ matrices which form a  a type 1 family with
$u=B$ and $\gamma_j^2=s_j$. The BCS model is obtained\cite{cambiaggio} as $\sum_j\eps_j \hat h_j$  for $\gamma_j^2=s_j=1/2$ and a replacement 
	$u=B\to 1/\mathrm{g}$, where $\mathrm{g}$ is the dimensionful BCS coupling constant.

Some blocks of the 1D Hubbard model characterized by a complete set of $u$-independent symmetry quantum numbers
are type 1 matrices, though most blocks are type $M>1$\cite{owusu1}. A similar typology can be developed for e.g. 
the 1D $XXZ$ Hamiltonian and other sectors of Gaudin and BCS models using the method of\cite{owusu1} for determining 
the type of parameter-dependent matrices. Interestingly, this implies that at least in some blocks there is an exact solution in terms of a single algebraic equation --- \eref{constraint1} or a similar equation for higher types\cite{owusu1} --- a vast simplification as compared to   Bethe's Ansatz.

One can also construct fermionic (bosonic) Hamiltonians\cite{shastry1} out of type 1 matrices as 
\beq 
\hat H=\sum_{mn} [H(u)]_{mn} a_m^\dagger a_n,  \label{fermi}
\eeq
 where $a_n$ are the usual fermionic (bosonic) destruction operators.
$[\hat H_1, \hat H_2]=0$ as long as the corresponding matrices $H_1(u)$ and $H_2(u)$ commute, i.e. belong to the same family.

\section{\label{lstat}Level statistics}

We have performed an extensive numerical study of level statistics of type 1 and higher type matrices\cite{hansen} for various choices of parameters. Almost in all cases the statistics is Poissonian  for  $N\gg 1$ with high accuracy, even when we deliberately attempt to adjust the parameters to get a different statistics.
Let us briefly describe the main results e.g.  for the level-spacing distribution.

It is convenient to redefine the parameter $u\to 1/x$ and replace $T+uV\to V+xT$. To get a proper large $N$ limit one has to make sure that the scaling of parameters $d_k, \eps_k, \gamma_k$ in \eref{H} and $x$ with $N$ is such that the eigenvalues
of $V$ and $xT$ scale in the same way for large $N$.  As discussed above, the matrix $V$ is arbitrary, so at $x=0$ one can have any admissible level statistics. Consider, for example, three representative cases: (a) $V$ is a random real symmetric matrix  with independent identically distributed matrix elements $V_{jk}$ for $j\le k$, (b) eigenvalues $d_k$ of $V$ are independent uniformly distributed random numbers, and (c) $d_k$ display level \textit{attraction}, 
$P(s)=as^\omega \exp(-b s^{1+\omega})$ with $-1<\omega<0$. The level-spacing distribution $P(s)$ for $V$ is Wigner-Dyson $P(s)=2ase^{-as^2}$ in (a) and Poissonian $P(s)=e^{-s}$ in (b). As soon as $V$ is chosen, $T$ is no longer arbitrary and we find that it has Poissonian $P(s)$ in all three cases for 
all choices of parameters $\gamma_k$ and $\eps_k$ we considered as long as $\eps_k$ and $d_k$ are uncorrelated.

Specifically, motivated in part by the BCS and Gaudin examples discussed in the previous section we took: (1) $\eps_k$ that are also eigenvalues of a random matrix and $\gamma_k=\mbox{const}$ independent of $k$, (2) same as (1) but with random uncorrelated $\gamma_k$ and (3) independent uniformly distributed $\eps_k$ and $\gamma_k=\mbox{const}$.  For all these choices  the level-spacing distribution for $T$ is very well approximated by Poissonian $P(s)=e^{-s}$, where $s$ is the level-spacing in units of the mean level-spacing. In case (a)
above the level statistics of $H(x)=V+x T$ at $x=0$ is Wigner-Dyson, but we find that it crosses over to Poisson at $|x|\approx 1/N$ and remains Poisson for larger $|x|$. Case (c) is analogous to (a) --  a crossover to Poisson behavior at $|x|\approx 1/N$. In case (b) the statistics is Poissonian for all $x$. Similar behavior is found in spectral rigidity. We conclude that
one can arrange for any statistics at a given value of the parameter $x=x_0$, but this becomes an isolated point in 
$N\to\infty$ limit, while for $x\ne x_0$  integrability as defined in Sect.~\ref{def} enforces Poisson statistics.

The only exception to Poisson statistics other than at an isolated value of $x$ we were able to identify is when parameters $d_k$ and $\eps_k$ are correlated, so that $d_k=f(\eps_k)$, where $f(\eps)$ is a smooth function of $\eps$ in $N\to\infty$ limit\cite{limit} and $\gamma_k=\mbox{const}$. This is the case in e.g. the BCS model where $d_k=\eps_k$ (see above). In such cases the statistics is distinctly non-Poissonian and, moreover, in case (a) above, for example, $P(s)$ crosses over at $x=O(N^0)\equiv O(1)$ from the Wigner-Dyson $P(s)=2as e^{-as^2}$ to a more repulsive distribution $P(s)\propto s^4$ for small $s$. The repulsion is  softened by randomizing  $\gamma_k$. More importantly, the statistics quickly becomes Poissonian when the correlation between $d_k$ and $\eps_k$ is destroyed, $d_k=f(\eps_k)(1+\eta_k)$, where $\eta_k$ are random. We find Poisson distribution already for
$\eta_k=O(1/N)$ at $x=O(1)$, see also \cite{relano} for a similar study of Gaudin model.   $d_k=f(\eps_k)$ define exceptional `surfaces' of certain measure zero in parameter space. This seems analogous to the harmonic oscillator exceptions to the Poisson distribution in classical integrable systems\cite{berry}.  There too one finds increased level repulsion for oscillators instead of Poisson $P(s)$.

Some of these numerical observations can be understood using perturbation theory. Energies to the first order in $x$ are given by the second equation in \re{H}, where we set $|\gamma_j|^2=1/N$ to achieve proper scaling for large $N$ as discussed above\cite{scaling}. We
have
\beg
 E_m(x)\approx d_{m} -\frac{x}{N}\sum_{j\neq
m} 
\left(\frac{d_{m} -d_{j} }{\varepsilon_{m}-\varepsilon_{j}}\right).
\label{pert}
\en
The first term comes from $V$, which we take to have Wigner-Dyson $P(s)$, the second -- from 
$T$, which is determined by the integrability condition \re{comm1} and whose level statistics we do not control. Let us estimate $x$ at which the two terms in \eref{pert} become comparable.  Without loss of generality we can take $d_k= O(N^0)=O(1)$ and we must also take $\eps_k= O(1)$ so
that $T$ and $xV$ scale in the same way for large $N$. Suppose $\eps_k$ are ordered as
$\eps_1<\eps_2<\dots <\eps_N$. When $d_k$ and $\eps_k$ are uncorrelated $d_m-d_j$ is  $O(1)$ when $j$ is close to $m$, i.e. when $(\eps_m-\eps_j)=O(1/N)$. The second term in \eref{pert} is then $x c_m \ln N$, where $c_m=O(1)$ is a random number only weakly correlated with $d_m$.

If we now order $d_m$, $c_m$ in general will not be ordered, i.e. if $d_{m+1}>d_m$ is the closest 
level to $d_m$ and therefore $(d_{m+1}-d_m)=O(1/N)$,   the corresponding difference 
$(c_{m+1}-c_m)=O(1)$. The contributions to level-spacings from the two terms in \eref{pert}
become comparable for $x=x_c\approx 1/(N\ln N)$. It makes sense that the second term introduces a trend towards Poisson distribution because it is a (nonlinear) superposition of $\eps_k$ and 
$d_k$ -- eigenvalues of two uncorrelated random matrices. Thus, we expect  a crossover from Wigner-Dyson to Poisson distribution at  $x=x_c$. 

This argument breaks down when $d_k=f(\eps_k)$, since in this case $(d_m-d_j)=O(1/N)$  when 
$(\eps_m-\eps_j)=O(1/N)$. The two terms in \eref{pert} become comparable only at $x=O(1)$
in agreement with the numerics for this case. Moreover, the second term   no longer trends towards Poisson statistics. Relaxing the correlation between
$d_k$ and $\eps_k$ with $d_k=f(\eps_k)(1+\eta_k)$ and going through the same argument, one expects a  crossover to Poisson statistics at  $x=O(1)$ for $\eta_k=O(1/N)$.

\section{Discussion}

A distinct feature of the notion of quantum integrability proposed in this paper is the  parameter ($u$) dependence. This is  in contrast to the classical notion that does not require any such dependence. We however find it necessary to be able to quantize the classical definition in a meaningful way. In this sense our definition is more demanding than its classical counterpart.

On the other hand, there is another distinction from the classical notion that makes our definition seem less stringent -- even a single nontrivial integral of motion linear in $u$ is sufficient to declare the model integrable. This was originally motivated  by the absence of a well-defined analog of the number of degrees of freedom  in quantum mechanics. There is evidence however that the presence of a single such integral is actually much more consequential in the quantum case. For example, for any $H(u)=T+uV$ with a single nontrivial commuting partner $I(u)=K+uW$\cite{how} one also finds numerous (about $N$ linearly independent) integrals of motion quadratic in $u$ that commute with both $H(u)$ and $I(u)$\cite{owusu2}. We note that a generic matrix of the form $A+uB$ has
no quadratic integrals other than its own square and a multiple of identity. 
One of the implications of this is that, for example, currents of higher order in the parameter in the
1D Hubbard and $XXZ$ models might follow from the linear ones, i.e. be in some sense trivial given 
the linear integral. These are however open questions that require further research.

One can also consider e.g. a situation when a Hamiltonian of the form $H(u)=T+uV$ or $T+uV+u^2W$ has no nontrivial integrals linear in $u$, but a number of quadratic  ones. Such systems exist and can also be classified and, at least some of them, explicitly parametrized.
Nevertheless, this  `higher order in the parameter' integrability seems less relevant as in most physical examples of parameter-dependent integrable models one is able to identify a parameter such that the Hamiltonian and at least one of the currents are linear in it.

 It is worth commenting on the relationship between our viewpoint and the usual set of 
 `beliefs' based  on model integrable systems. Two related points emerge (i) the belief that the number of constants of motion is $\sim L \propto \log({\cal N})$ where $L$ is the number of sites and ${\cal N}$ the size of the total Hilbert space and  (ii) the so called `rule of three', i.e.  the belief that any many body lattice model in 1-dimension  with a fixed number of particles reveals its integrability only in the three particle sector\cite{sutherland-3}, since the one particle sector and two particle sectors have as many constants of motion as the particle  number (total energy and momentum).
 
 With regard to (i), we distinguish between the much larger ${\cal N}$ and $N$ of this work.   By a process of block diagonalizing the Hamiltonian operator into different sectors, one arrives at a direct sum representation of the full Hamiltonian. Each  sub block is `irreducible' in the sense that the space-time and internal space (parameter-independent)
 symmetries have been extracted out, and our considerations revolve around such  sub blocks with a smaller and variable dimension $N$. Our point is that any such  sub block must be special in the sense discussed here. 
 Stitching back the irreducible blocks to reconstruct the full Hamiltonian matrix requires a detailed knowledge of the symmetries used in the first place. While possible in principle,  we regard this process as  of secondary importance as compared to the one undertaken here, namely the characterization of the sub blocks themselves.  

 With regard to (ii) our studies of  two typical examples give some insight into this  question.
 Firstly, the Gaudin magnets show that the two and higher particle number sectors
yield matrices that have non linear conservations laws in addition to the linear ones discussed here.
 Secondly, we can study the fermionic  representation of type 1 matrices, \eref{fermi}. Here $n=L=N$, the sector $n_e=1$ is isomorphic to type 1 matrix family \re{H}, while other sectors are much larger matrices of high types $M={N\choose n_e}-N+1$\cite{except}, i.e. they are apparently distinct and much more complicated integrable matrix families. With the hindsight of 
\eref{fermi} it is evident that they all are  different manifestations of the same type 1 model and are in some sense equivalent, but can this be formulated generally so that all such similar  matrix integrable models are naturally grouped together and recognized to be related to each other as different representations of the same underlying structure?   Further work is needed to obtain  clarity on these questions. Recent work \cite{haile-new} throws light on  further symmetries in higher particle number sectors that are linear in the parameter $u$, and explores their non linear relationship to the operators  in \eref{fermi}.

Other open questions include: a general construction of type $M>3$ integrable families, analytical results for crossings in types $M>1$ and for level statistics in all types, the relationship between
the exact solution for types $M>1$ through a single algebraic equation and Bethe's Ansatz.

\begin{acknowledgments}

E.A.Y. is grateful to UCSC Physics Department, where part of this research was conducted, and especially to Sriram Shastry for hospitality. The work at UCSC  was  supported in part by DOE under Grant No. FG02-06ER46319.
E.A.Y. also acknowledges financial support by the David and Lucille Packard Foundation and the National Science
Foundation under Award No. NSF-DMR-0547769.  We thank our collaborators Haile Owusu and Daniel Hansen for helpful discussions.

\end{acknowledgments}


\begin{thebibliography}{99}
\bibitem{comment-finite}  Imagine that we are  given a specific Hermitean
  $6 \times 6$ matrix, can we determine  if it is the descendent of a quantum integrable model in a particular number-symmetry subsector? An explicit example might be from the  numerical study   in Ref.[\onlinecite{heilman}], where a number of matrices are found by restricting the Hubbard model to a particular number sector with all space time  symmetries
 factored out. We provide a systematic method of looking at such questions in this work.
\bibitem{localization} For example, the Anderson localization of electrons in a random potential is often viewed as a transition from  Poisson (localized) to GOE (extended) statistics.
 In this context it is not  obvious if or how the concept of  quantum  integrability  plays a role.
\bibitem{deg}  More precisely it reads $T_{km}^i(d_k^j-d_m^j)=T_{km}^j(d_k^i-d_m^i)$. If $d_k^i=d_m^i$ for some, but not all $i$, $S_{km}$ is defined through the non-degenerate eigenvalues. If $d_k^i=d_m^i$ for all $i$,  matrices $V^i$ share a common $2\times 2$ identity block. Then with a $u$-independent unitary transform one can go to a basis where $T_{km}^i=0$ for all $i$. In either case \eref{T} still holds.
\bibitem{degenerate} Degenerate $d_k$ are regarded as level crossings at 
$u=\infty$. That such definition is necessary is  seen e.g. by redefining the parameter $u\to 1/x$ and multiplying the matrices by $x$. We have the same commuting family now linear in $x$, but with the crossing moved to $x=0$. Note also that we cannot fix the order of
$d_k$ without loss of generality, because we have already chose our indexing so that $\eps_k$ are ordered.
\bibitem{yang}  If we set $\gamma_i=1$, this construction is a simple extension of  the  permutation operator for $SU(N)$ in Yang's well known  work \cite{yang1}   into matrix space, but seems to be new for the general case $\gamma_i \neq 1$.
\bibitem{limit}  To obtain $f(\eps)$ in this limit, we scale $\eps_k$ and $d_k$ with $N$ so that they fill finite intervals when $N\to\infty$.
\bibitem{scaling} Equivalently, one can redefine $x\to x/N$ as in the BCS model for which $x=\mathrm{g}=\lam \delta$, where $\lam$ is the dimensionless coupling and $\delta \propto 1/N$ is the mean level spacing in $\eps_k$.
\bibitem{how} Such examples are available analytically (type $N-2$ matrices) or can be generated numerically
by solving \eref{comm2}, see\cite{owusu1} for details.
\bibitem{except} Individual sectors do not have additional sector-specific $u$-dependent integrals. This is the case for the   fermion \re{fermi} and Gaudin models, but not for the Hubbard model where there are numerous such sector-specific  integrals\cite{owusu1,owusu2}.
\end{thebibliography}
\end{document}